\documentclass[iop]{emulateapj}
\usepackage{graphicx,epsfig,fancyhdr,rotating,amsmath,natbib}
\usepackage{txfonts}
\usepackage{epsfig,rotate}
\usepackage{natbib}

\newcommand{\degree}{\ensuremath{^\circ}}
\shorttitle{Spectroscopic Observations of a Coronal Loop}
\shortauthors{Gupta, Tripathi and Mason}
\begin{document}
\title{Spectroscopic Observations of a Coronal Loop: 
Basic Physical Plasma Parameters Along the Full Loop Length}
\author{G.~R. Gupta, Durgesh Tripathi}
\affil{Inter-University Centre for Astronomy and Astrophysics, Post Bag-4, Ganeshkhind, Pune 411007, India}
\email{girjesh@iucaa.ernet.in, durgesh@iucaa.ernet.in}
\author{Helen E. Mason}
\affil{Department of Applied Mathematics and Theoretical Physics, University of Cambridge, Wilberforce Road,
 Cambridge CB3 0WA, UK}
\email{h.e.mason@damtp.cam.ac.uk}
\begin{abstract}

Coronal loops are the basic structures of the solar transition region and corona. 
The understanding of physical mechanism behind the loop heating, plasma flows, and filling are still considered a major challenge in the solar physics.
The mechanism(s) should be able to supply mass to the corona from the chromosphere and able to heat the plasma over 1~MK within the small
distance of few hundred km from the chromosphere to the corona. This problem makes coronal loops an interesting target for detailed study.
In this study, we focus on spectroscopic observations of a coronal loop, observed in its full length, in various spectral lines as 
recorded by the Extreme-ultraviolet Imaging Spectrometer (EIS)
on-board Hinode. We derive physical plasma parameters such as electron density, temperature, pressure, column depth,
and filling factors along the loop length from one foot-point to the another. The  obtained parameters are used to infer
whether the observed coronal loop is over-dense or under-dense with respect to gravitational stratification of the solar
atmosphere. These new measurements of physical plasma parameters, from one foot-point to another,
provide important constraints on the modeling of the mass and energy balance in the coronal loops.

\end{abstract}


\section{Introduction}\label{intro}

The problem of solar coronal heating has been one of the most stubborn problems in the field of astrophysics. In spite of major 
developments in observational capabilities and theoretical modeling, the problem remains to be solved in its full glory. For an 
excellent extensive review see e.g., \citet{2006SoPh..234...41K}. 

Loops are considered to be the building blocks of the corona. Current observations suggest that there are different types of loops 
seen at different temperatures. Broadly, these loops could be divided into three groups based on their characteristic temperatures, 
namely fan loops [$<$~1 MK], warm loops [seen at around 1~MK], and hot loops mainly seen in the core of active regions in high 
temperature lines [3-5~MK]. However, there are numerous cooler closed loop structures observed 
in spectral lines such as \ion{O}{5} and \ion{Si}{7} i.e. below 1 MK and they may not be classified as fan loops \citep[see e.g.,][]{1997SoPh..175..511B}. In addition 
to the well defined loop structures it has been found that there is a significant amount of diffuse emission in and around the active 
regions \citep[see e.g][]{2006SoPh..234...41K}. In a recent study,  \cite{2014ApJ...795...76S} have shown that  these regions have a 
characteristic temperature of about 2~MK. In fact, it has also been found that the distinguishable loops have only about 20-30\% larger 
intensities than the foreground/background diffuse emission \citep{2003A&A...406.1089D,2012ApJ...753...35V}. Although, the exact 
quantity of the diffuse emission varies depending on the contamination of the instrumental stray light.  Therefore, for any 
coronal heating theories, it is important to explain all these different kinds of emission from active regions. Although, the loops are 
just above 20-30\% in brightness from the diffuse regions, they provide one of the best targets of opportunity to study the mass and 
energy circulation from transition region to the corona.

In order to understand the emission of these different kinds of loop structures, it is most important to study their properties using 
spectroscopy in addition to imaging. On the one hand, spectroscopic techniques provide the most accurate physical parameters such 
as density, temperature and flows. On the other hand imaging can provide the time evolution of the structures. Since we are interested in 
understanding the emission from quiescent active regions, a study of the time evolution is extremely important to rule out any effects of solar activity 
phenomena such as flares, micro-flares etc on the observed loop structures.

Warm loops, which emanate mostly at the periphery of active regions have been most easily detectable with instruments 
such as the Coronal Diagnostic Spectrometer \citep[CDS, ][]{1995SoPh..162..233H} on-board Solar and Heliospheric Observatory (SOHO), 
Transition Region and Coronal Explorer \citep[TRACE, ][]{1999SoPh..187..229H}, Solar Terrestrial Relations Observatory (STEREO), Extreme-Ultraviolet 
Imaging Spectrometer \citep[EIS, ][]{2007SoPh..243...19C} on-board Hinode \citep{2007SoPh..243....3K} and the Solar Dynamics 
Observatory (SDO). Therefore, warm loops have been  most extensively studied and physical plasma parameters have been measured 
using CDS/SOHO \citep[see e.g.,][]{2001ApJ...556..896S,2003A&A...406.1089D,2007ApJ...655..598C} and  EIS/Hinode 
\citep[see e.g.][ and thereon citations]{2008ApJ...686L.131W, 2009ApJ...694.1256T, 2011ApJ...730...85B, 2012SoPh..276..113S, 2013ApJ...772L..19B}. 

In this paper, we report for the first time, the analysis of a complete loop using a spectrometer, although extensive information exist in
the literature on loop segments \citep[see e.g.][]{2003A&A...406.1089D, 2005A&A...439..351U, 2008ApJ...686L.131W, 2009ApJ...694.1256T, 2012SoPh..276..113S}. 
The loop is observed with EIS/Hinode and the Atmospheric Imaging Assembly \citep[AIA, ][]{2012SoPh..275...17L} 
on board SDO. This observation provides us with an excellent 
opportunity to measure the physical parameters of the loop from one foot-point to the another, providing unique observational constrains to the 
modeling of coronal loops. The rest of the paper is structured as follows. In section~\ref{obs} we describe the observations, followed by the 
data analysis and results in section~\ref{analysis}. We summarize our results and draw some conclusions in section~\ref{conclusion}.

%%%%------------------------------------------------------------
\section{Observations}\label{obs}
%%%%------------------------------------------------------------

%%%%------------------------------------------------------------
\begin{figure*}[htbp]
\centering
\includegraphics[width=14cm]{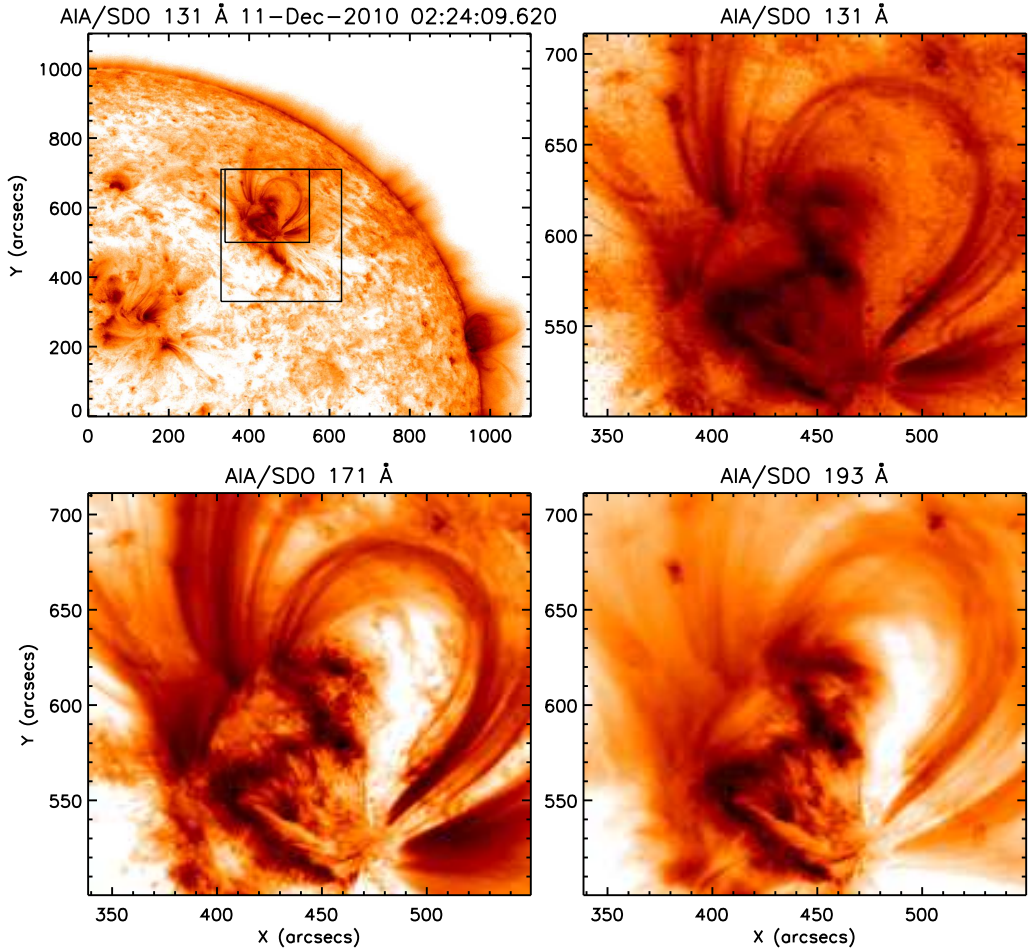}
\caption{Top left image: a portion of the Sun's disk showing the active region AR 11131 in the 131~{\AA} passband. The bigger box indicates the region which was 
rastered by EIS/Hinode. The smaller box is shown in subsequent images taken in AIA 131~{\AA}, 171~{\AA}, and 193~{\AA} passbands as labeled.}
\label{fig:context}
\end{figure*}
%%%%------------------------------------------------------------
%%%%------------------------------------------------------------
\begin{figure*}[htbp]
\centering
\includegraphics[width=14cm]{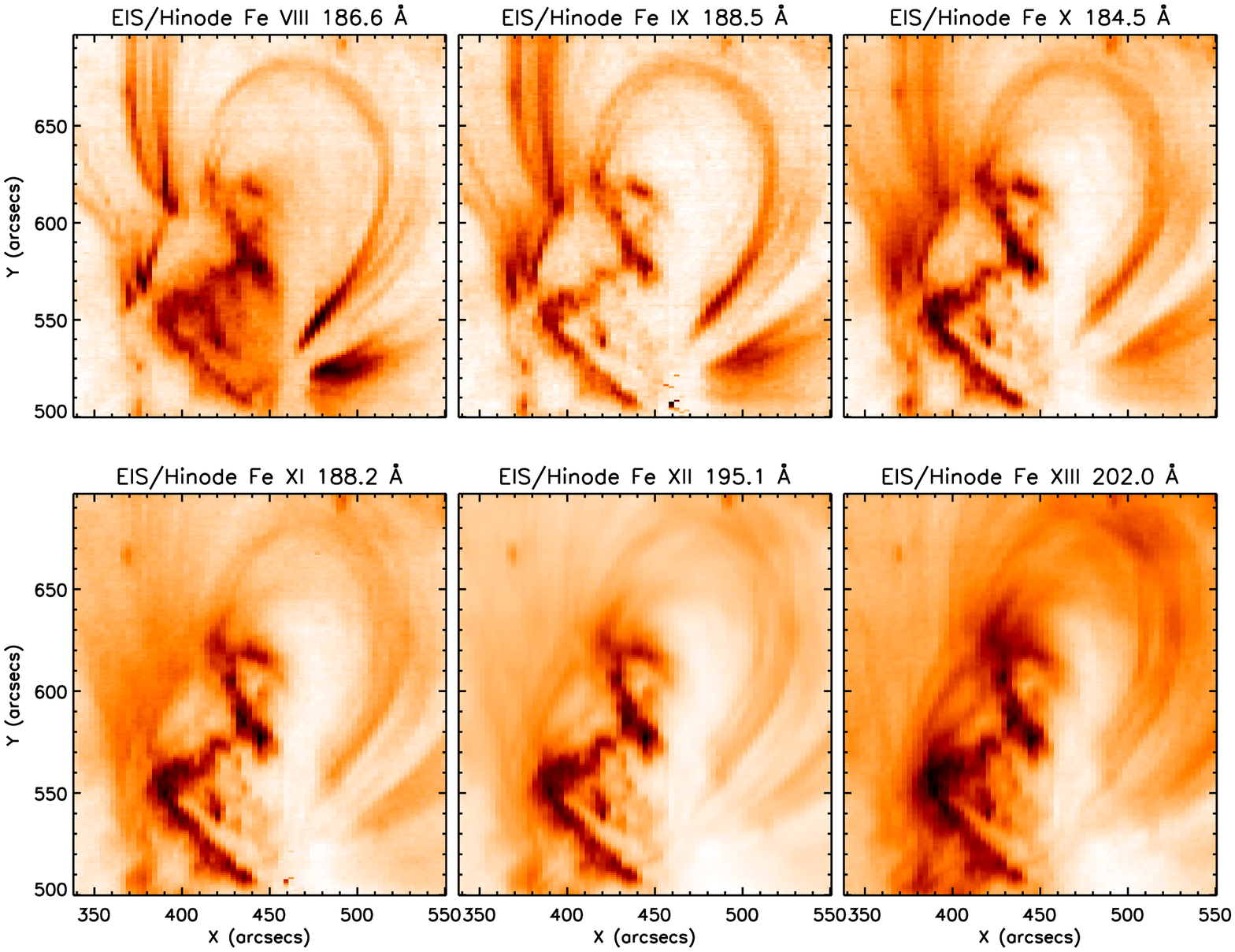}
\caption{Monochromatic intensity maps of an active region AR 11131  obtained in different wavelengths using EIS/Hinode (as labeled). 
Maps correspond to the upper half of the box shown in top right panel of Figure~\ref{fig:context}.}
\label{fig:context2}
\end{figure*}
%%%%------------------------------------------------------------

A coronal loop from one foot-point to another, was detected in an active region, \textit{AR 11131} observed on 
2010 December 11 with EIS \citep[][]{2007SoPh..243...19C} on-board Hinode 
\citep{2007SoPh..243....3K}.  Observations were carried out with the 2\arcsec\ slit with an exposure time of 35~s and a sparse 
raster of the region was created with a step size of 3\arcsec. The raster scan started at 01:55~UT and was completed 
at 02:56~UT covering a field of view of $384\arcsec \times 300\arcsec$. 
%\textbf{This dataset was previously analyzed by \citet{2012ApJ...754L...4T} to study plasma flows in a warm coronal loop.}
We followed standard procedures for analyzing the EIS data using
software available in \textsl{Solar Software} \citep[SSW;][]{1998SoPh..182..497F} to calibrate and correct for the slit tilt and orbital drift. 
The spatial offset in the solar-Y direction
at different wavelengths was corrected by co-aligning the images at each wavelengths.  EIS \ion{Fe}{8}~186.6~\AA\ and 
\ion{Mg}{7}~278.4~\AA\ spectral lines were chosen to  co-align the spectral images from short and long wavelength windows as the
peak formation temperature of both the lines are similar, thus, revealing the similar structures.
Images were aligned using a cross-correlation technique and are accurate up to the 1\arcsec\ difference in solar-Y direction, however,
 no spatial offset in the solar-X direction in two wavelength windows was observed, as the raster step size was 3\arcsec\ in this observation.

Figure~\ref{fig:context} provides the context image of the observed coronal loop from  AIA/SDO  and EIS/Hinode. The top left panel
displays a part of the Sun's disk covering the active region as observed in AIA 131~{\AA} passband. The emission in the
131~{\AA} passband has contributions mainly from \ion{Fe}{8}, \ion{Fe}{21} and  \ion{Fe}{23} spectral lines. 
However, for the quiescent loops,  \ion{Fe}{8}  dominates the emission. For more information on dominant emission in various AIA channels in 
different regions see \citet{2010A&A...521A..21O} and  \citet{2011A&A...535A..46D}. The bigger box on the top left panel indicates the EIS raster
field of view whereas the smaller box is the part of active region chosen for the current analysis. The chosen region
clearly shows the complete coronal loop from one foot-point to another. The remaining three images corresponds to the smaller box 
shown in the top left panel. The top right panel shows the loop in the AIA 131~{\AA} passband whereas the bottom panels show the AIA 
171 ~{\AA} (bottom left) and AIA 193~{\AA} (bottom right) passbands. The loop structure can be identified in all the three passbands, 
however, the structure changes in different passbands.  The AIA 131~{\AA} passband shows a well defined loop structure with very little 
background/foreground emission, whereas in AIA 171~{\AA} passband shows an enhancement in background/foreground emission. In the 
AIA 193~{\AA} passband, it becomes difficult to disentangle  the loop structure from the foreground/background emission. 

\begin{figure*}[htbp]
\centering
\includegraphics[width=15cm]{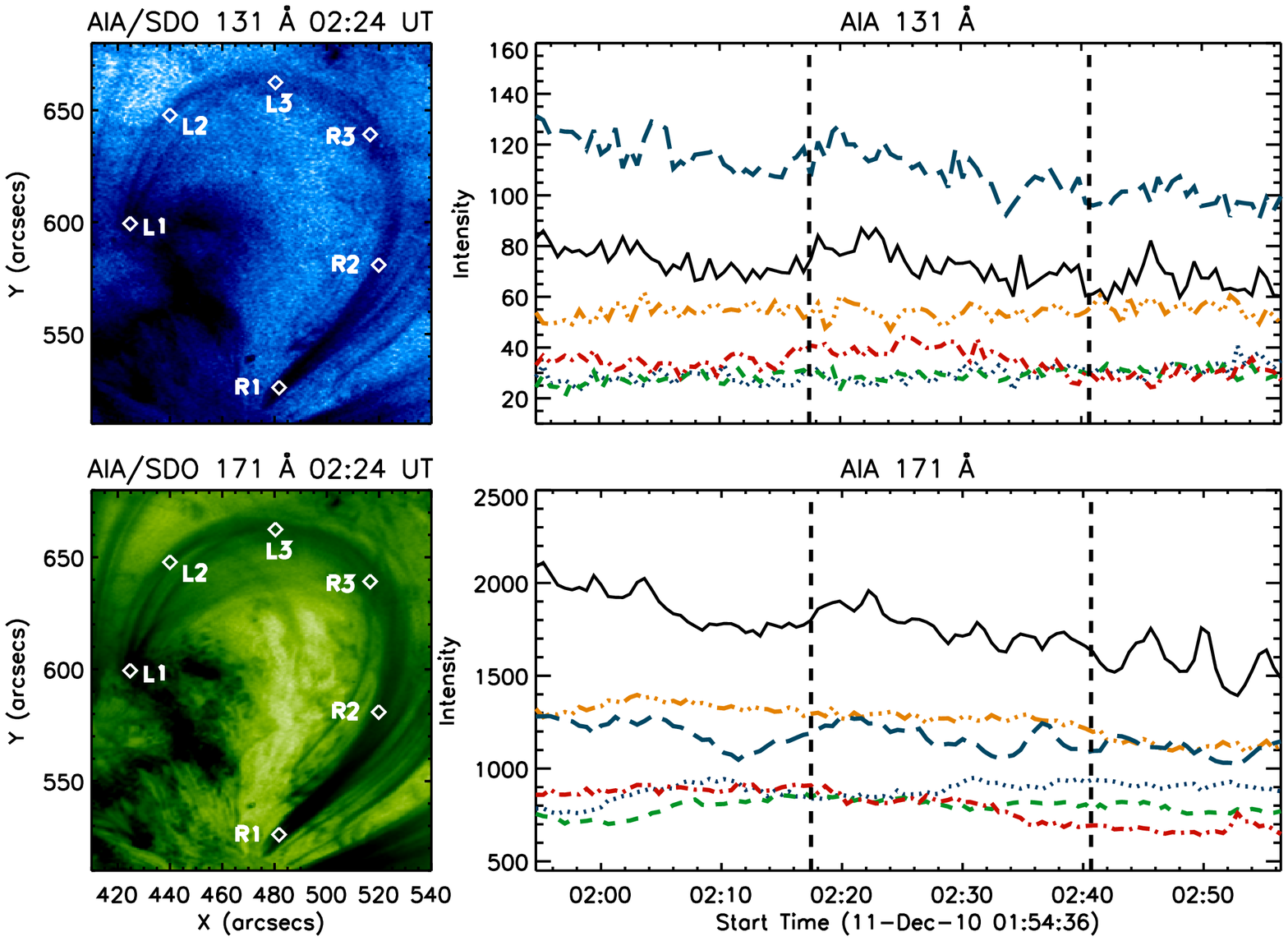}
\caption{AIA 131~{\AA}, and 171~{\AA} light curves at different locations of the coronal loop during the time interval of EIS raster scan. 
Plotted curves corresponds to L1 (solid line), L2 (dotted), L3 (dashed), R3 (dash dot), R2 (dash dot dot dot), and R1 (long dash) locations along the loop.
Two vertical dashed lines in each panel indicate the time taken to scan the observed coronal loop using EIS.}
\label{fig:aia_lc}
\end{figure*}

As noted by  \citet{2012ApJ...754L...4T}, this loop is quiescent and quasi-static, and does not show any flaring and/or micro-flaring 
events before or after the acquired EIS raster. In order to reconfirm this, we performed a quantitative study as follows. We obtained 
light curves at different locations along the loop during the time interval of EIS raster scan for AIA 131~{\AA}, and 171~{\AA} emission. The 
obtained light curves, shown in Figure~\ref{fig:aia_lc}, reveal that there are no significant changes occurring during the span of the raster scan. 
However, a slight decrease in intensity can be observed at few locations along with some small enhancements in the light curves. From these 
curves, it is plausible to conclude that within the given time scale of raster scan the observed loop can be considered as quiescent and 
quasi-static. This makes it a suitable observation for the study of the basic plasma parameters along the full loop length. 

%%%%------------------------------------------------------------
%%%%------------------------------------------------------------

%%%%------------------------------------------------------------
\section{Data analysis and Results} \label{analysis}
%%%%------------------------------------------------------------

Figure~\ref{fig:context2} displays EIS intensity maps of the coronal loop structure obtained in different spectral lines 
(\ion{Fe}{8}, \ion{Fe}{9}, \ion{Fe}{10}, \ion{Fe}{11}, \ion{Fe}{12}, and \ion{Fe}{13}) corresponding to the smaller box shown in 
Figure~\ref{fig:context}. The loop structure visible in the EIS \ion{Fe}{9} image is very similar to that seen in the AIA 171~\AA\ passband. 
The loop structure becomes less distinguishable with increasing temperature. The loop can still be seen in \ion{Fe}{11}, where the loop 
intensity is still higher than the background. The loop gets completely merged within the background/foreground in the images obtained 
from \ion{Fe}{12} and \ion{Fe}{13}, showing similar characteristics to those discussed by \citet{2009ApJ...694.1256T} and \citet{ 2010ApJ...719..576G}.

The aim of this study is to derive physical plasma parameters such as electron density, temperature, and filling factors along the loop 
structure. For the purpose of deriving the electron density, we have used the line pair of \ion{Mg}{7} $\lambda 278.39/\lambda 280.75$.
The \ion{Mg}{7}~278.39~{\AA} spectral line is blended with  \ion{Si}{7}~278.44~{\AA} \citep{2007PASJ...59S.857Y}. However, using the density insensitive ratio of 
\ion{Si}{7}~278.44~{\AA} with 275.35~{\AA} with a fixed branching ratio of 0.32, the contribution of \ion{Si}{7}~278.44~{\AA} can be removed from  
\ion{Mg}{7}~278.39~{\AA} \citep{2007PASJ...59S.727Y}.

%%%%------------------------------------------------------------
\begin{figure*}[htbp]
\centering
\includegraphics[width=14cm]{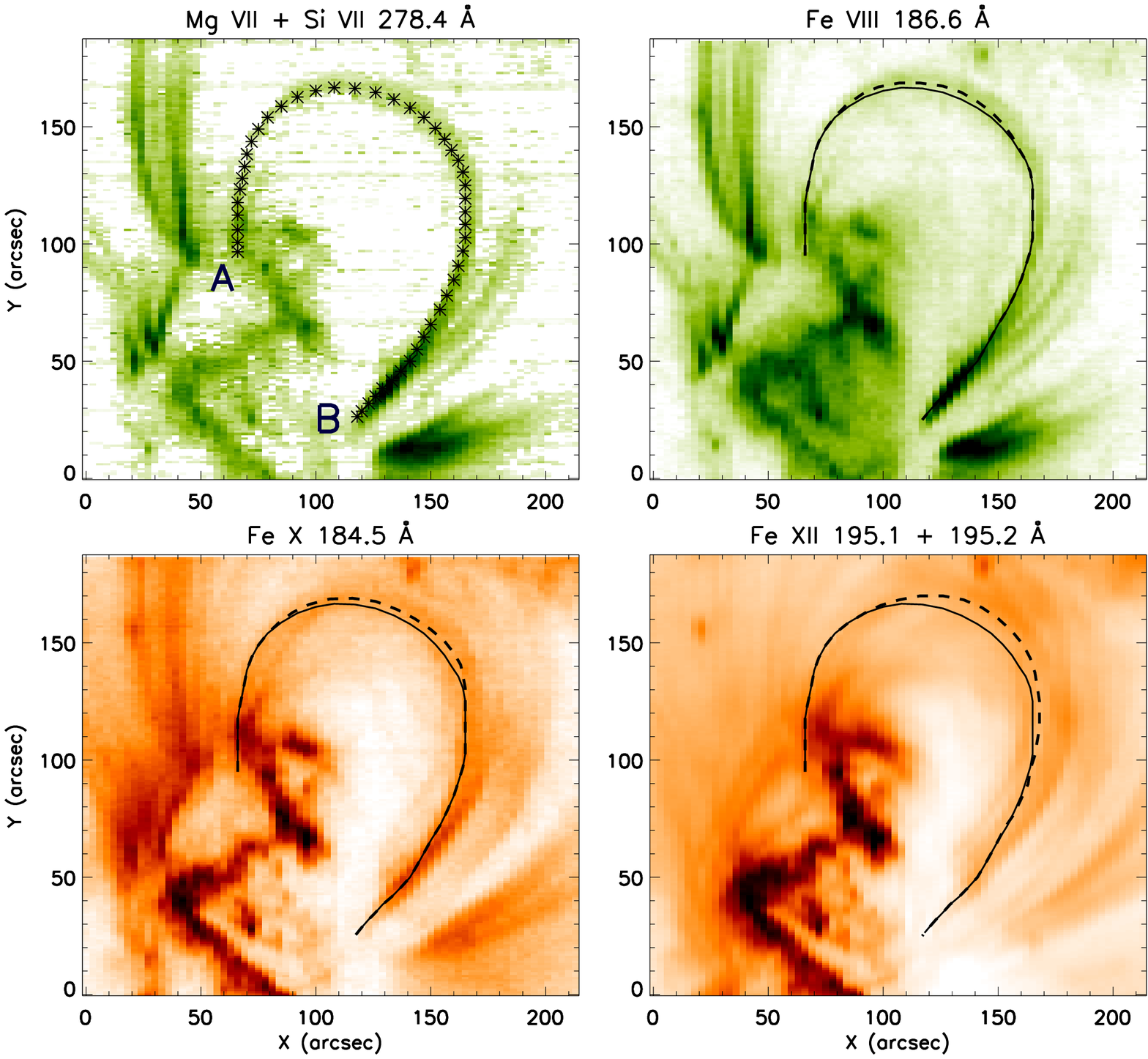}
\caption{The traced coronal loop from left foot-point A to right foot-point B as seen in the Mg~{\sc vii} 278.39~\AA\ (blended with Si~{\sc vii} 278.44~\AA ) spectral line (top left).
The loop is also traced manually in  Fe~{\sc viii}~186.6~\AA\ (top right), Fe~{\sc x}~184.5~\AA\ (bottom left), and Fe~{\sc xii}~195.12~\AA\ (blended with Fe~{\sc xii}~195.18~\AA ) 
(bottom right) spectral lines marked with dashed lines. Traced loop structure from Mg~{\sc vii} 278.39~\AA\ is also over-plotted with a continuous lines in all the other panels. }
\label{fig:loop_tracing}
\end{figure*}
%%%%------------------------------------------------------------

Figure~\ref{fig:loop_tracing} displays intensity maps showing the coronal loop obtained in Mg~{\sc vii}~278.39~\AA\  (blended with Si~{\sc vii} 
278.44~\AA ), Fe~{\sc viii}~186.6~\AA , Fe~{\sc x}~184.5~\AA , and Fe~{\sc xii}~195.12~\AA~(blended with Fe~{\sc xii}~195.18~\AA) spectral 
lines after fitting the line profiles with single Gaussian function.  The over-plotted asterisk signs in the 
Mg~{\sc vii}~278.39~\AA\ intensity map indicate the manual tracing of the loop. The loop is also traced in other spectral lines and over-
plotted with dashed lines. For comparison, we over-plotted the traced loop co-ordinates of Mg~{\sc vii}~278.39~\AA\ with a continuous line on 
top of the intensity maps obtained from higher temperature spectral lines. These images reveal that with increasing temperature, the loop top 
extends to higher altitude. The difference between the apex region is not so different between \ion{Mg}{7} and \ion{Fe}{10}, but the difference 
becomes significant between \ion{Mg}{7} and \ion{Fe}{12} and is more than 7\arcsec . Since great care has been taken with the co-alignment, 
we believe that this is real and not due to any instrumental effects (images at different wavelengths are aligned with an accuracy of 
about 1\arcsec). Therefore, it is plausible to conclude that the structures seen in \ion{Fe}{12} are different to those seen in \ion{Mg }{7}.

For the present analysis, we have considered the loop structure observed in \ion{Mg}{7}. In order to improve the signal to noise ratio, we 
binned over three pixels in the Y-direction for each of the tracked points along the loop. All the resultant profiles are fitted 
with Gaussian to extract the line parameters along the loop. The contribution of \ion{Si}{7}~278.44~{\AA} has been subtracted from the 
\ion{Mg}{7}~278.39~{\AA} spectral line using the constant branching ratio of \ion{Si}{7}~278.44~{\AA} with \ion{Si}{7}~275.3~{\AA}. 

\begin{figure}[htbp]
\centering
\includegraphics[width=0.45\textwidth]{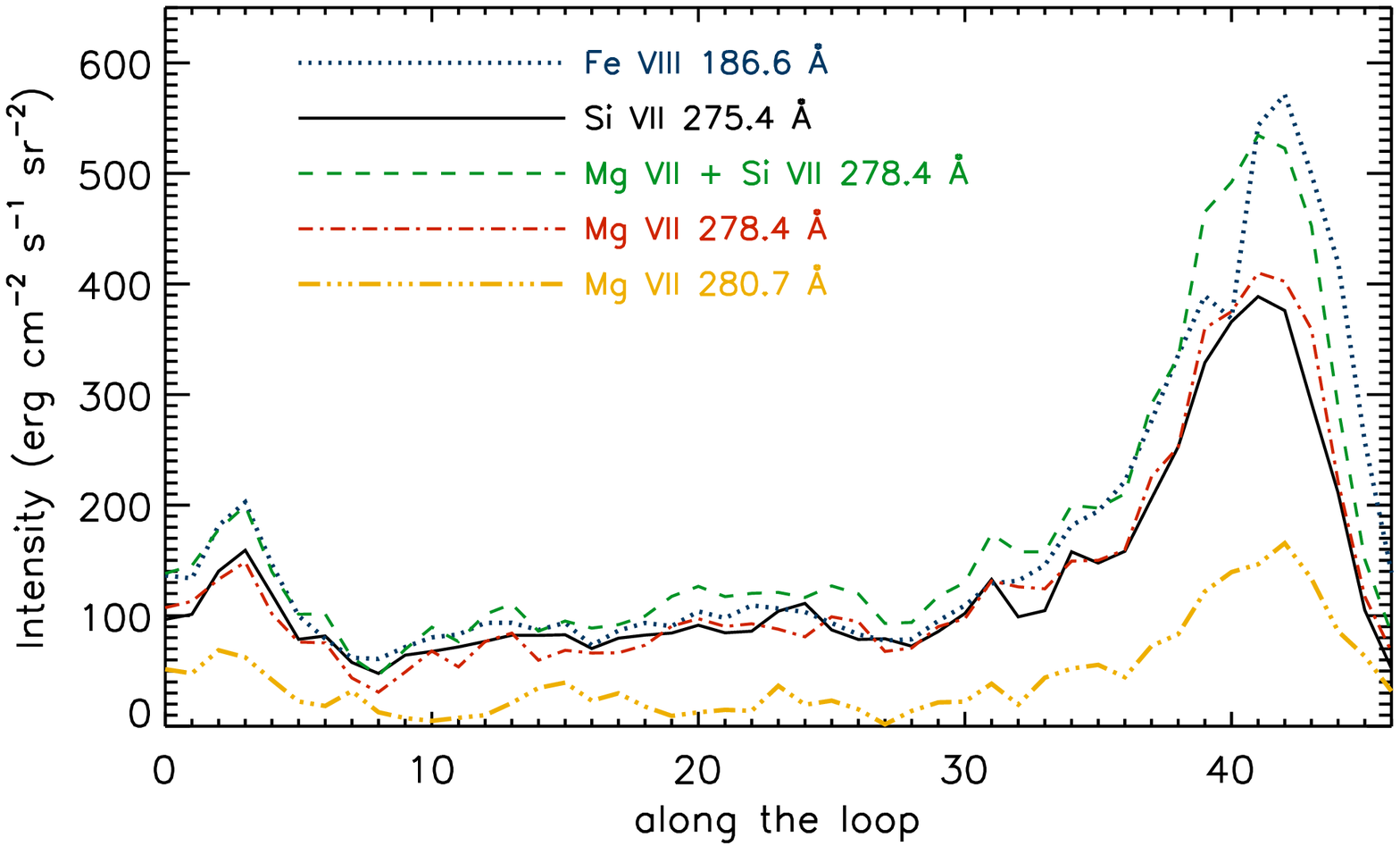}
\caption{The variation of intensity along the coronal loop from left foot-point A to right foot-point B as measured in different spectral lines from  EIS/Hinode.}
\label{fig:loop_int}
\end{figure}

Figure~\ref{fig:loop_int} shows the variation of intensity along the loop in various spectral lines as labeled in the plot . These intensities were 
corrected for the sensitivity degradation with time of the EIS instrument as per the scheme suggested by \cite{2013A&A...555A..47D}. As evident 
from the figure, the observed loop is brightest  in the \ion{Fe}{8}~186.60~{\AA},  whereas it is barely visible in the \ion{Mg}{7}~280.75~{\AA} spectral 
line, except the two foot-points. The plot also suggest that the intensity at the foot-point B is a factor of 4-5 larger than that at foot-point A. This is very 
likely caused by foot-points of various loop structures aligned along the line of sight, as can also be seen from the AIA images.

%%%%------------------------------------------------------------
%%%%------------------------------------------------------------

%%%%------------------------------------------------------------
\begin{figure}[htbp]
\centering
\includegraphics[width=0.45\textwidth]{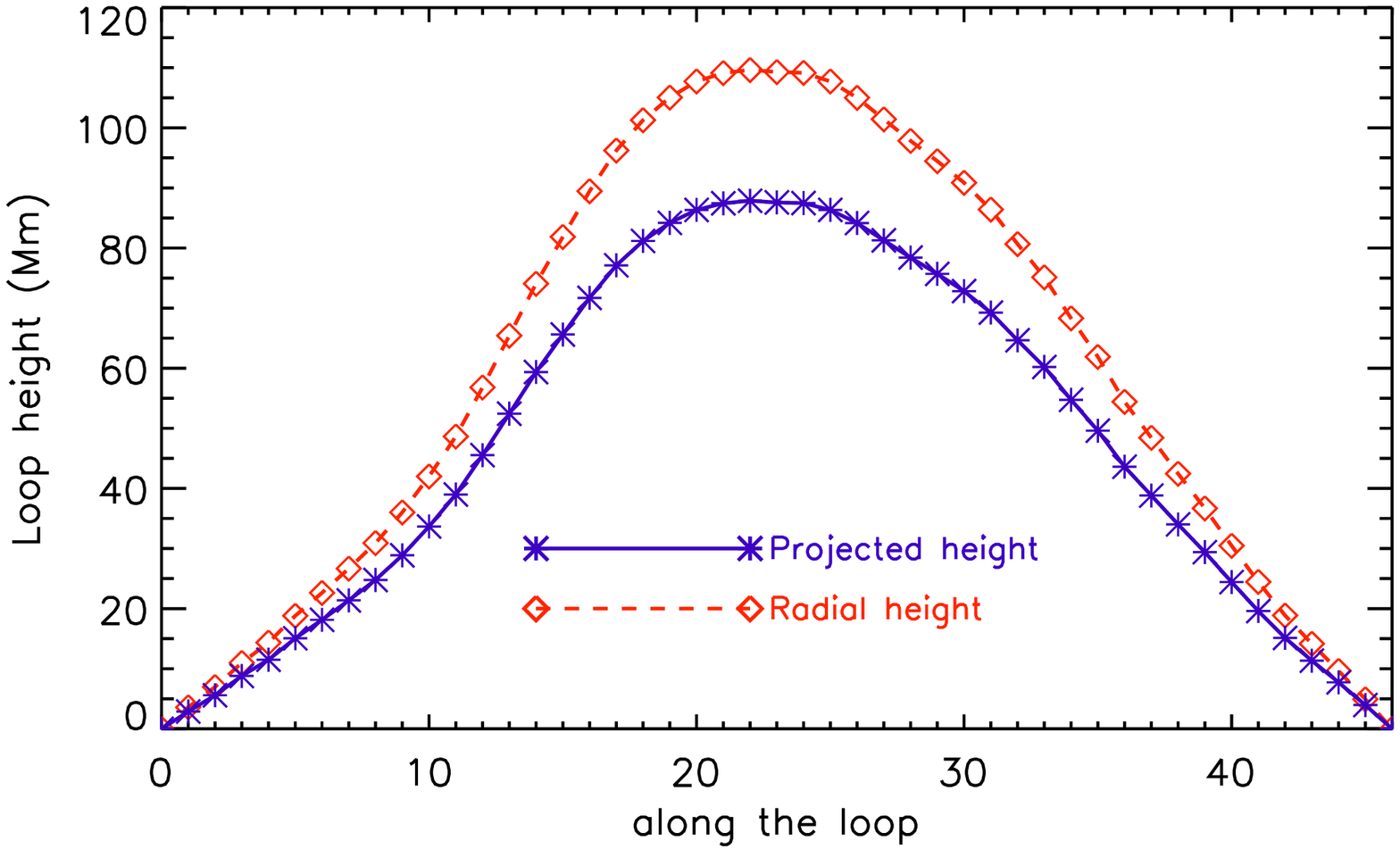}
\caption{The height of traced coronal loop from left foot-point A to right foot-point B as measured
with respect to the loop foot-points and obtained  from Mg~{\sc vii}~278.39~\AA\ intensity map. }
\label{fig:loop_height}
\end{figure}
%%%%------------------------------------------------------------

In order to estimate the height of loop with respect to the foot-points, we use the following method. We first draw a straight line connecting 
the two foot-points. The lines normal to the line connecting the two foot-points will provide the projected height extent of the loop coordinates 
with respect to the footprints in the solar-XY plane. The estimated height extent of loop points are plotted in Figure~\ref{fig:loop_height} and marked with asterisks. 
The projected loop top extends up to the height of $\approx 85$~Mm (position 22--23). 

In order to get the exact height of the loop, we need to correct for the projection effect. 
This would require information of the inclination angle of the loop plane with respect to the solar-XY plane. 
This is a non-trivial exercise, as we don't have observations from any other vantage points.
Unfortunately, the separation angle between the two STEREO 
spacecrafts was so large at the time of observation that only STEREO-A could see the loop. Additionally, the angle between SDO and 
STEREO-A was also too large to perform any stereoscopic studies. 
Therefore for the current analysis, we assumed that the coronal loop would be  extending radially outward, and found that 
the loop would be inclined towards the observer at an angle of about $\sim36\degree$ (for an average loop location 
at $X\approx 450\arcsec$, $Y\approx 580\arcsec$ and solar radius of $\approx 980\arcsec$ ) with respect to the solar-XY plane.
In this case, the loop top would extend up to the height of $\approx 110$~Mm (marked with diamonds in Figure~\ref{fig:loop_height}, position 22--23). 

%%%%------------------------------------------------------------
\begin{figure*}[htbp]
\centering
\includegraphics[width=16cm]{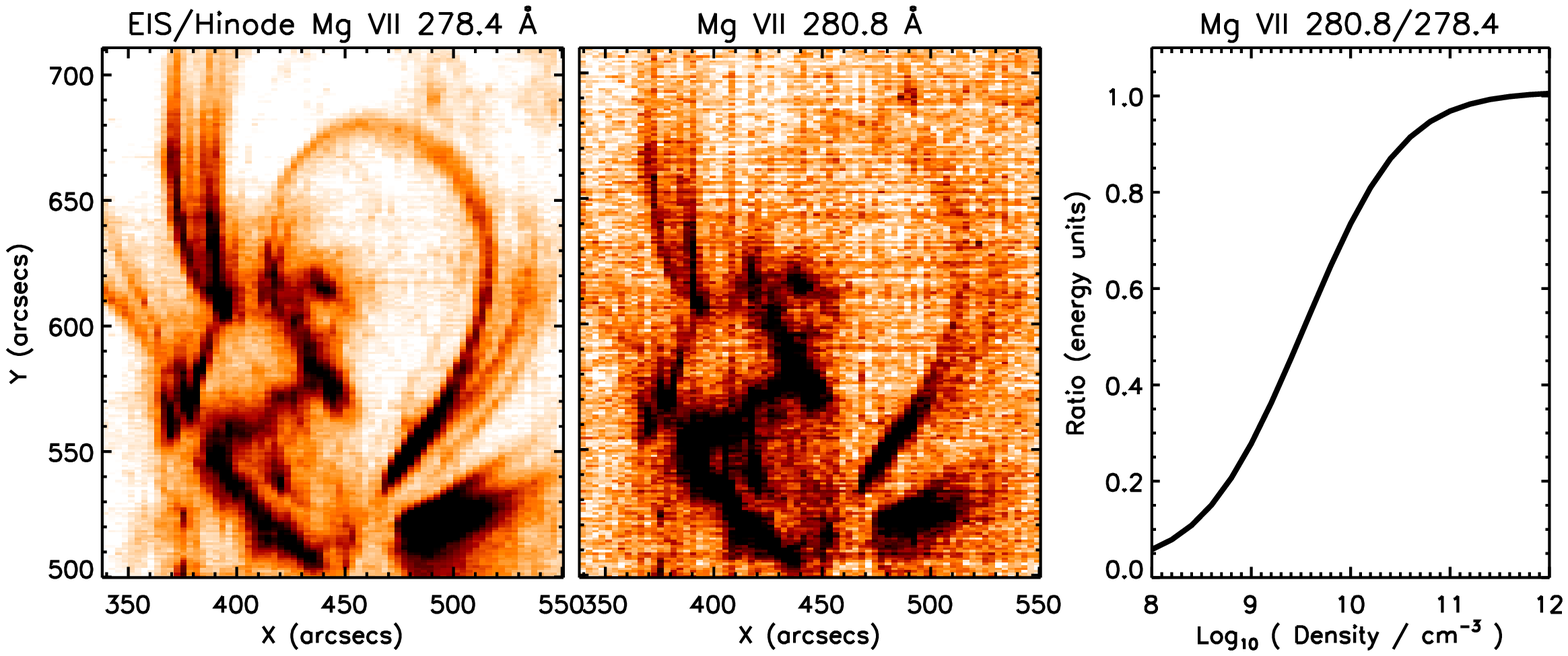}
\caption{Intensity maps of coronal loop obtained in Mg~{\sc vii} 278.39~\AA\ (left panel) and 280.75~\AA\ (middle panel) spectral lines.
Intensity ratio of two spectral lines is sensitive to electron density and variation is obtained from CHIANTI version 7.1 (right panel).}
\label{fig:mg7}
\end{figure*}
%%%%------------------------------------------------------------

To estimate the basic plasma parameters along the loop, we employ methods used by \citet{2010A&A...518A..42T} and  described in 
subsequent subsections.

%%%%------------------------------------------------------------
\subsection{Density along the loop length}\label{dens}
%%%%------------------------------------------------------------

The electron density of a stellar atmosphere can be derived by taking the intensity ratio of two emission lines of the same ion having different 
dependence on the electron density \citep[e.g. ][]{1994A&ARv...6..123M,2009A&A...495..587Y}. Densities obtained from this method, do not
depend on the emitting volume, elemental abundances, ionization state of the plasma, and purely depend on the atomic population 
processes within the ion and provides the mean electron density along the line of sight.

EIS provides access to a number of density sensitive line ratios formed at different temperature ranges \citep{2007PASJ...59S.857Y}. We 
measured electron number density along the full loop length using the density sensitive line ratios from the different ions using the CHIANTI 
version 7.1 \citep{1997A&AS..125..149D,2012ApJ...744...99L}. 
We choose a pair of  \ion{Mg}{7} $\lambda 278.39/\lambda 280.75$  density sensitive lines to obtain the electron density at various points 
along the loop length. The loop is clearly visible in the Mg~{\sc vii} 278.39~\AA\ spectral line along its full length. A distinguishable structure can also be seen
 in Mg~{\sc vii}~280.75~\AA\, however, it is barely visible towards the loop top, indicating lower electron density towards the loop top
in comparison to the foot-points (see Figure~\ref{fig:mg7}). Therefore as mentioned earlier, in order to obtain a good signal, we binned over 
3-pixels in Y-direction at all the chosen points along the loop so as to be able to perform the Gaussian fit and obtain the intensities in both the 
spectral lines.

\begin{figure}[htbp]
\centering
\includegraphics[width=0.45\textwidth]{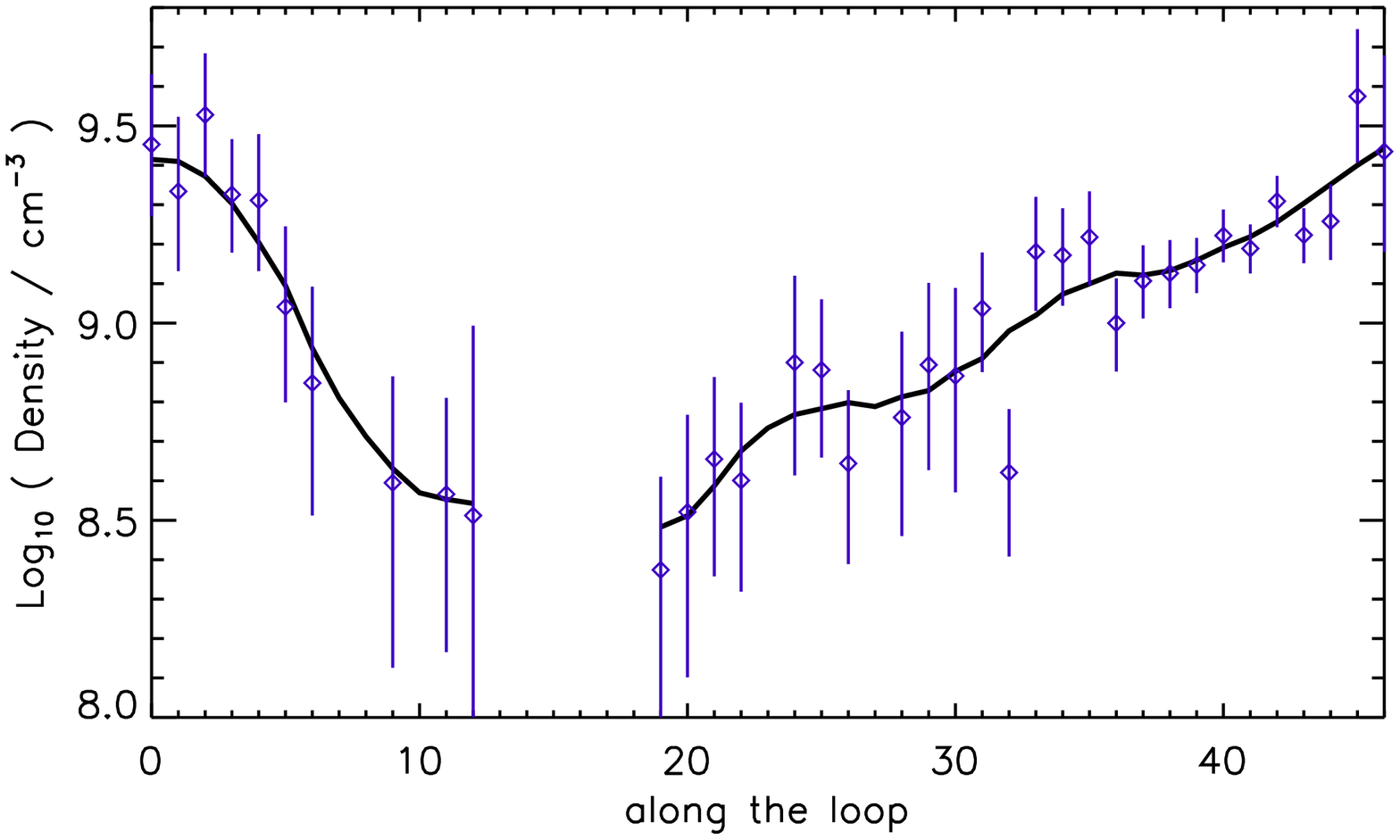}
\caption{Density measured along coronal loop from left foot-point A to right foot-point B using Mg~{\sc vii}  $\lambda 280.75/\lambda 278.39$ 
line ratio. 5-point running average of data points is also over-plotted with continuous line.}
\label{fig:dens_mg7}
\end{figure}

Figure~\ref{fig:dens_mg7} shows the density variation along the loop as measured using the \ion{Mg}{7} line 
pair. We also obtained densities after performing background/foreground subtraction. However, densities before and after 
background subtraction do not show any significant change mainly due to very low background/foreground emission as can be seen from 
Figure~\ref{fig:mg7}. The density plot shows that the density drops off with the height of the 
loop. The number densities at both the foot-points are similar and have values $\approx 10^{9.4}$~cm$^{-3}$. The density decreases upon moving 
towards the loop-top and attains the lowest value of about $\approx 10^{8.5}$~cm$^{-3}$ (see Figure~\ref{fig:dens_mg7}, position 15--16). 
Moving towards the loop top, the signal strength of  \ion{Mg}{7} 280.75~\AA\ goes down resulting in larger error bars. 
As mentioned earlier, towards the loop top there is hardly any signal, thus the number density estimate towards the loop top was not possible. 
We will use this electron density obtained from \ion{Mg}{7} line ratio for the purpose of further analysis.

%%%%------------------------------------------------------------
%%%%------------------------------------------------------------

%%%%------------------------------------------------------------
\subsection{Temperature along the loop length}
%%%%------------------------------------------------------------

Basic physical plasma parameters along the loop length such as temperature and filling factors, can be derived using spectroscopic techniques as follows:

The expression for the optically thin emission line intensity can be given as

\begin{equation}
 I = 0.83~Ab(z)~\int G(T_e,N_e)~N^2_edh
\label{eq:int}
\end{equation}

\begin{figure}[htbp]
\centering
\includegraphics[width=0.45\textwidth]{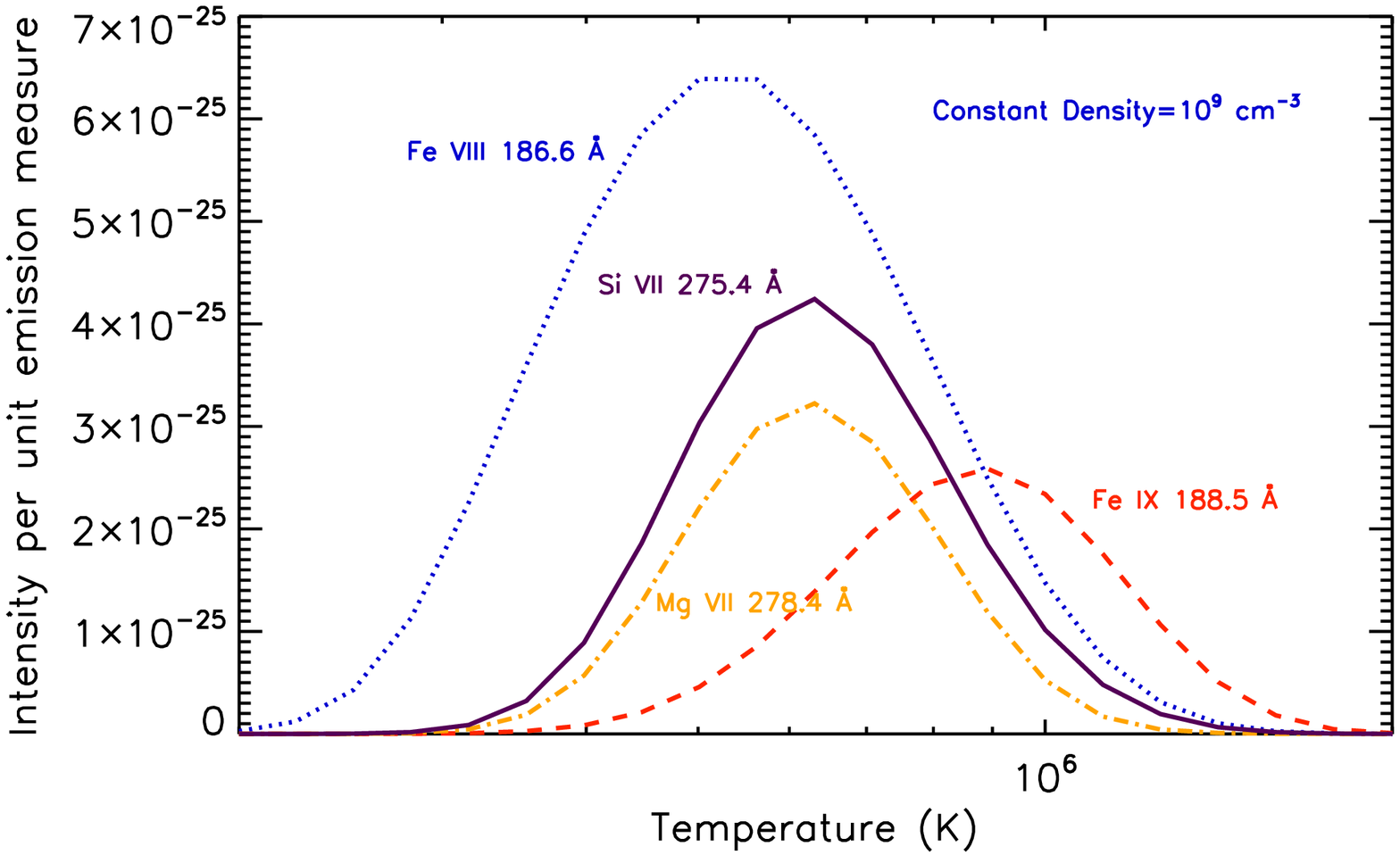}
\caption{Contribution function of various spectral lines as labeled and used in this study.} \label{fig:goft}
\end{figure}

Where $Ab(z)$ is the elemental abundances, $N_e$ is the electron number density, $h$ is column depth of emitting plasma along the line of sight. 
The factor 0.83 is the ratio of protons to free electrons which is a constant for temperatures above $10^5$ K.  $G(T_e,N_e)$ is
the contribution function which contains all the relevant atomic parameters for each spectral line and is defined as

\begin{equation}
 G(T_e, N_e) =  \frac{hc}{4\pi \lambda_{i,j}} \frac{A_{ji}}{N_e} \frac{N_j (X^{+m})}{N(X^{+m})} \frac{N(X^{+m})}{N(X)}   
\label{eq:g}
\end{equation}

where $i$ and $j$ are the lower and upper levels, $A_{ji}$ is spontaneous transition probability, $\frac{N_j (X^{+m})}{N(X^{+m})}$ is the population of level $j$ relative to the total
$N(X^{+m})$ number density of ion $X^{+m}$ and is a function of electron temperature and density, $ \frac{N(X^{+m})}{N(X)}$ is the ionization fraction and is predominantly a function
of temperature. All the calculations were performed using the CHIANTI version 7.1 \citep{2012ApJ...744...99L,2013ApJ...763...86L} available in SSW distribution and using the 
coronal abundances of \citet{1992PhyS...46..202F}. Contribution function $G(T_e, N_e)$ of various spectral lines are plotted in Figure~\ref{fig:goft}.

The column emission measure (EM) can be defined as:

\begin{equation}
 EM= \int N_{e}^{2}dh
\label{eq:emd}
\end{equation}

Thus, the column emission measure (EM)  can then also be written as

\begin{equation}
 EM= \frac{I_{obs}}{0.83~Ab(z)~G(T_e,N_e)}
\label{eq:em}
\end{equation}

%%%%------------------------------------------------------------
%%%%------------------------------------------------------------

\begin{figure*}[htbp]
\centering
\includegraphics[width=16cm]{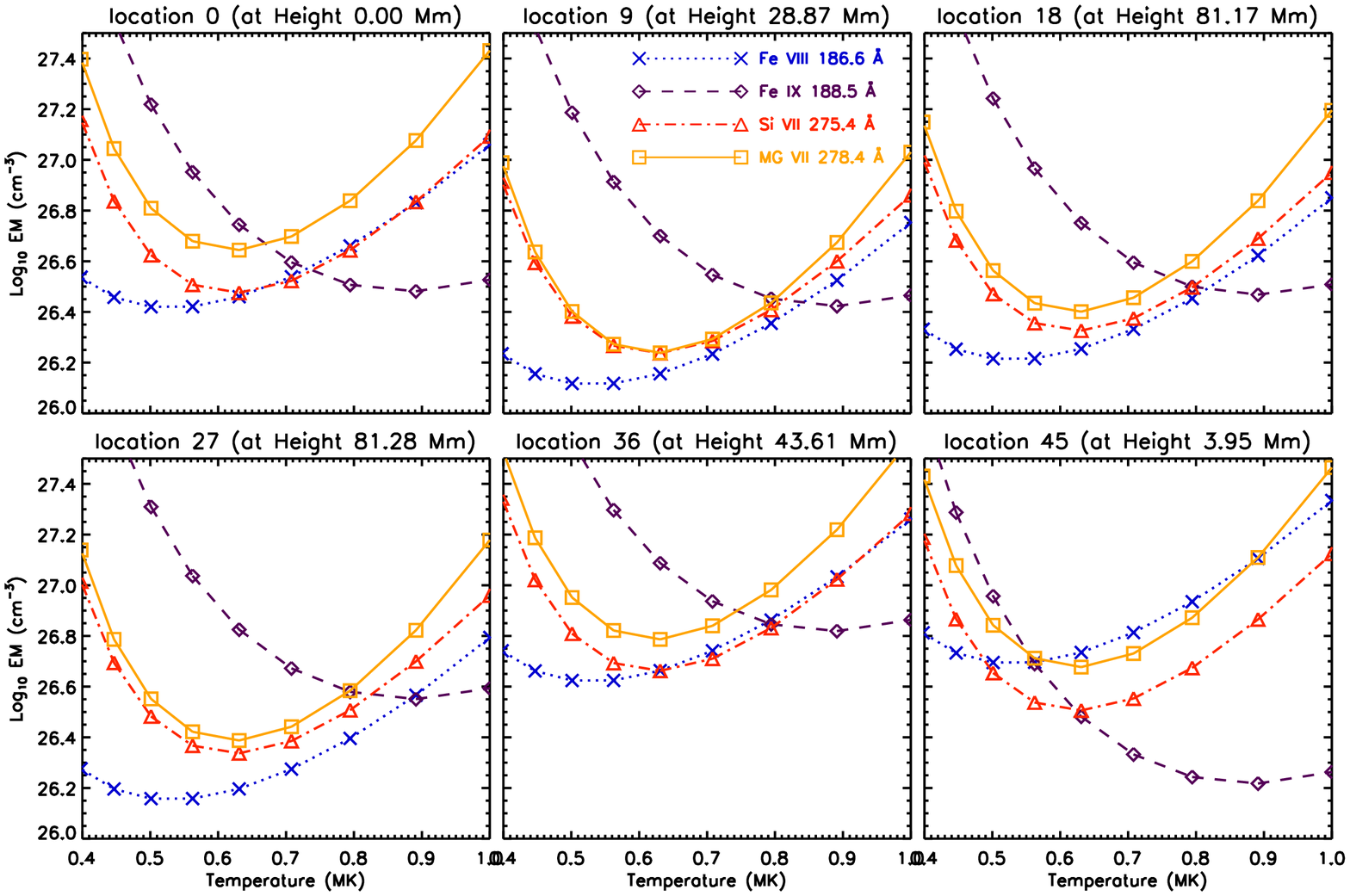}
\caption{EM loci curves of several spectral lines obtained at various locations along the coronal loop.}
\label{fig:em_loci}
\end{figure*}

An estimate of the electron temperature of plasma can be obtained using emission lines from ions with different ionization stages. As 
contribution functions of spectral lines are highly dependent on temperatures, the observed intensity can be converted in to temperature
upon analyzing a broad range of ion species. An effective technique used for estimating the electron temperature is called the EM-loci 
method \citep[e.g., ][]{1987MNRAS.225..903J,2002A&A...385..968D}. In this method, the ratios of observed intensities of different spectral 
lines with their corresponding contribution functions and abundances (i.e., EM as defined in Equation~\ref{eq:em}) are plotted as a function of 
temperature. If the plasma is iso-thermal along the line-of-sight then all of the curves would cross at a single location indicating a single 
temperature. In this study, we used density insensitive emission lines Fe~{\sc viii}~186.61~\AA , Fe~{\sc ix}~188.5~\AA , 
Si~{\sc vii}~275.35~\AA , and Mg~{\sc vii}~278.39~\AA\ to obtain the EM loci curves. We obtained EM loci curves at all the locations along 
the loop from point A to point B as labeled in Figure~\ref{fig:loop_tracing}. However, we have plotted only a few of them in Figure~\ref{fig:em_loci}
to show the trend in the temperature variation. The EM curves obtained from these spectral lines cross (or have point of closest approach) at 
almost the same temperature, suggesting that the loop is nearly iso-thermal along its length. A similar result was found in 
previous studies using CDS and EIS \citep[see e.g.,][]{2003A&A...406.1089D,2008ApJ...686L.131W,2009ApJ...694.1256T}. 

\begin{figure}[htbp]
\centering
\includegraphics[width=0.45\textwidth]{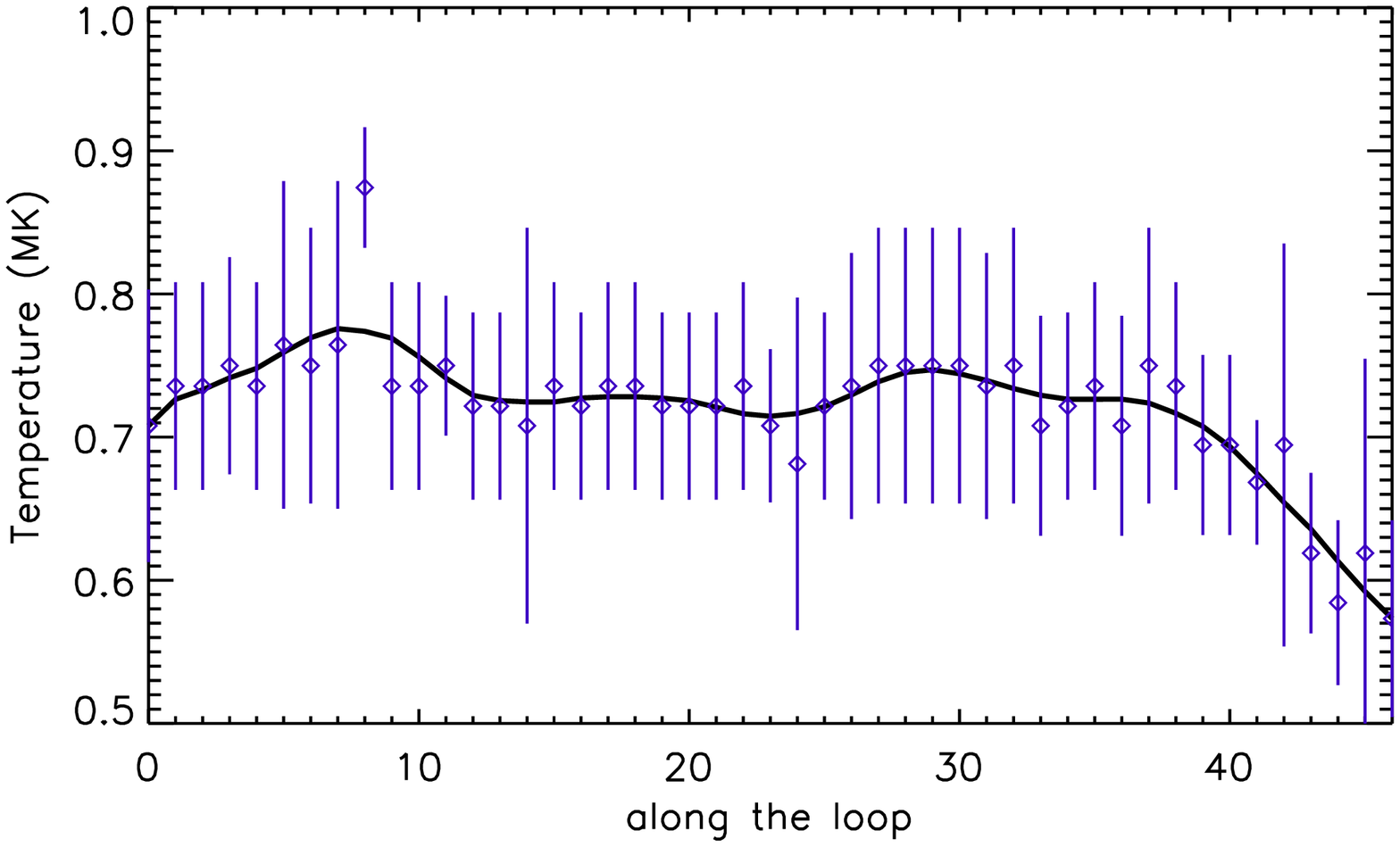}
\caption{Temperature variation along the coronal loop from left foot-point A to right foot-point B as obtained from the EM loci curves.
5-point running average of data points is also over-plotted with continuous line.}
\label{fig:temp_loci}
\end{figure}

\begin{figure}[htbp]
\centering
\includegraphics[width=0.45\textwidth]{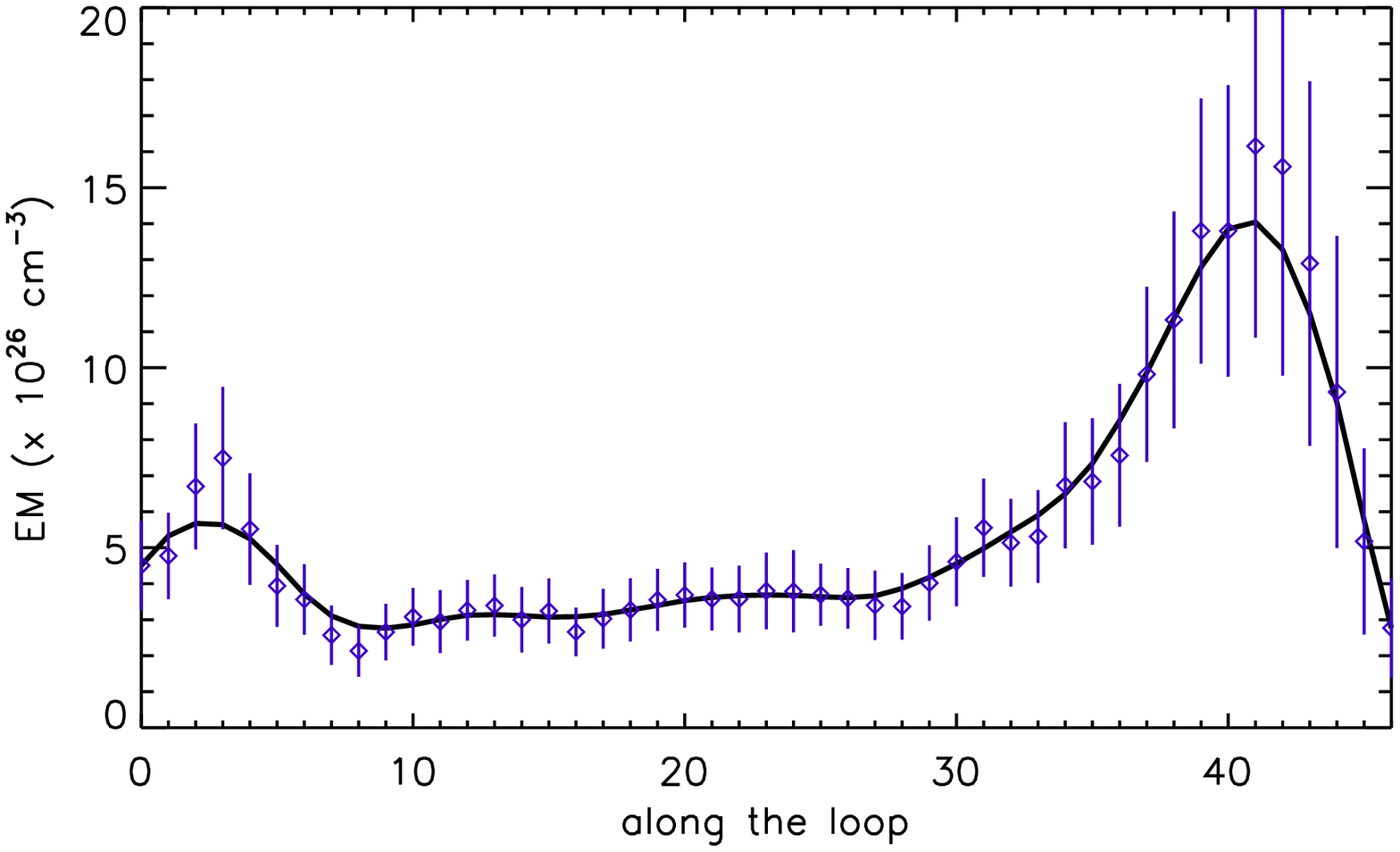}
\caption{Average EM variation along the coronal loop from left foot-point A to right foot-point B as obtained from the EM loci curves.
5-point running average of data points is also over-plotted with continuous line.}
\label{fig:em_loop}
\end{figure}

There are in total 6 intersection points for all the four spectral lines at each locations. We took mean of these points as the electron temperature 
at that location and the standard deviation of these points to be an error-bar on the temperature at that location. The resultant temperature 
variation along the loop as obtained from EM-loci curve method is shown in Figure~\ref{fig:temp_loci}. The figure indicates that the temperature 
is almost constant along the loop within the obtained error-bar, however, a  slight decrease is observed near the right foot-point B. 
We also obtained the average EM distribution along the loop length in a similar way and this is shown in Figure~\ref{fig:em_loop}.
The average EM variation along the loop is similar to that of the intensity variation along the loop as found in Figure~\ref{fig:loop_int}.
The EM-loci method has been used  to study temperature structure across various points along the partial loop segment by various
authors using EIS data \citep[see e.g.][]{2008ApJ...686L.131W, 2009ApJ...694.1256T, 2012SoPh..276..113S}. \cite{2012SoPh..276..113S}
in particular found almost a constant temperature of $\log\,T=6.2$ along a loop segment using similar method. In the present study,
the average temperature along the loop length is found to be about 0.73~MK ($\log\,T=5.86$) as obtained from the spectral lines in which loops were clearly
visible. However, we would also like to point out that \cite{2013ApJ...765L..46K} found temperature gradient along the different off-limb loop structures 
using temperature sensitive emission line ratio method.

%%%%------------------------------------------------------------
%%%%------------------------------------------------------------

%%%%------------------------------------------------------------
%%%%------------------------------------------------------------

%%%%------------------------------------------------------------
%%%%------------------------------------------------------------

\begin{figure}[htbp]
\centering
\includegraphics[width=0.45\textwidth]{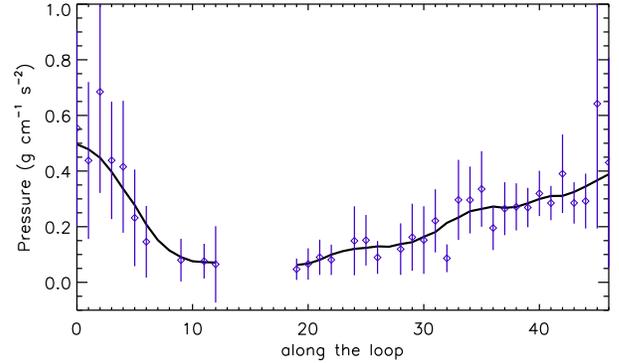}
\caption{Pressure variation along the coronal loop from left foot-point A to right foot-point B as obtained from the ideal gas equation.
5-point running average of data points is also over-plotted with continuous line.}
\label{fig:pressure}
\end{figure}

As noted in Figures~\ref{fig:dens_mg7} and \ref{fig:temp_loci}, the density decreases whereas temperature remains almost steady upon moving from the loop foot-points to 
towards the loop top. Thus, we obtained the thermal pressure along the loop using the ideal gas equation $p=2N_e k_b T_e$ \citep[for the coronal pressure, ion pressure 
($p_i$) and electron pressure ($p_e$) are added together, ][]{2005psci.book.....A} and plotted in Figure~\ref{fig:pressure}. We see that pressure at left foot-point A 
($p\approx0.5$ g cm$^{-1}$ s$^{-2}$) is slightly higher than that at right foot-point B ( $p\approx0.4$ g cm$^{-1}$ s$^{-2}$). 
However, the measured pressure difference at two foot-points are rather very small and may arise due to some small evolutionary changes in the loop while raster scan.

%%%%------------------------------------------------------------
%%%%------------------------------------------------------------

\subsection{Filling factor along the loop length}

A knowledge of the coronal filling factor ($\phi$) which is a ratio of volume radiating in X-rays and EUV to the total volume, is an important parameter in order to 
understand the coronal heating \citep{1993SoPh..147..263C}. This provides information on the observed structures whether they are resolved or unresolved 
with the given instrument. Almost all the one dimensional hydrodynamic models require the coronal structures observed with current day instrumentation to be 
multi-stranded in order to explain the observations \citep{2006SoPh..234...41K}. Therefore, a clear understanding of filling factors would provide crucial inputs to 
the theoretical modeling of coronal structures and thereby would help us understand the coronal heating problem better. We estimate the filling factor along the 
coronal loop using the process similar to that of \citet{2010A&A...518A..42T} and described below. 

As noted earlier, emission measure (EM) can be estimated from the two different methods (Eqs.~\ref{eq:em} and \ref{eq:emd}).
Thus, plasma filling factor ($\phi$) can be obtained from  \citep{1997ApJ...478..799C,2010A&A...518A..42T},

\begin{equation}
 \phi=\frac{EM}{N_e^2 h}
\label{eq:phi}
\end{equation}
where $EM$ is measured using Equation~\ref{eq:em} and N$_e$ is measured in section~\ref{dens}.

\begin{figure}[htbp]
\centering
\includegraphics[width=0.45\textwidth]{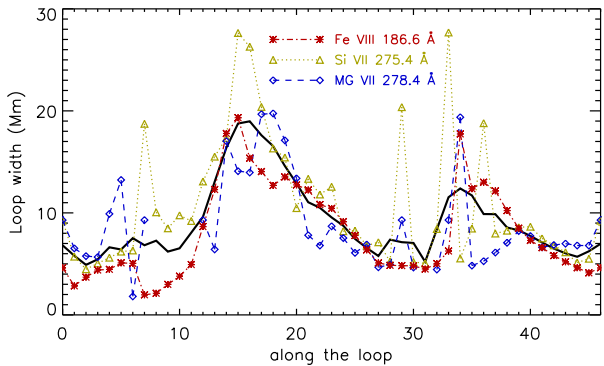}
\caption{Geometric width of coronal loop  from left foot-point A to right foot-point B as seen
in different spectral lines. The dark continuous  line provides average width obtained from different spectral lines along the loop length.}
\label{fig:cdepth}
\end{figure}

The most important parameter, therefore, is to estimate column depth $h$ from observations in order to estimate the filling factors.
In this case, the column depth at each position was measured assuming that loop has a cylindrical geometry. Thus, the projected loop diameter 
(geometric width of the loop)  will be equal to column depth ($h$). We estimate the loop diameter to be equal to full width half maxima (FWHM) 
obtained after fitting a Gaussian function across the loop points. The estimated column depth which is the geometric width of the loop along 
the loop length is plotted in Figure~\ref{fig:cdepth}. We obtained the geometric width of the loop in several spectral lines in which the loop is 
clearly visible. The plot shows many fluctuations in the loop width, which can also be seen in the intensity maps. The average loop width is 
almost constant as we move from left foot-point A, and increases towards the loop top and decreases further after attaining a maximum width at 
the loop top (see Figure~\ref{fig:cdepth}, position 15-16). Towards the right loop foot-point B, the loop width again increases. This increase in the loop width is due to the 
fact that there are many more structures emanating from that location which are aligned along the loop as can be see in the images 
recorded by AIA/SDO as shown in Fig.~\ref{fig:context}. We will use this average  loop width along the loop length for the further analysis in
this study. It is also interesting to note that loop width has  increased by a factor of about two  between tracked points  10 and  16. 
This may indicate that the observed loop may not have a cylindrical geometry,
and has different orientation angle along the loop with respect to line of sight. However, we will keep assuming 
the cylindrical geometry of the loop for further analysis in this study. 

%%%%------------------------------------------------------------
%%%%------------------------------------------------------------

\begin{figure}[htbp]
\centering
\includegraphics[width=0.45\textwidth]{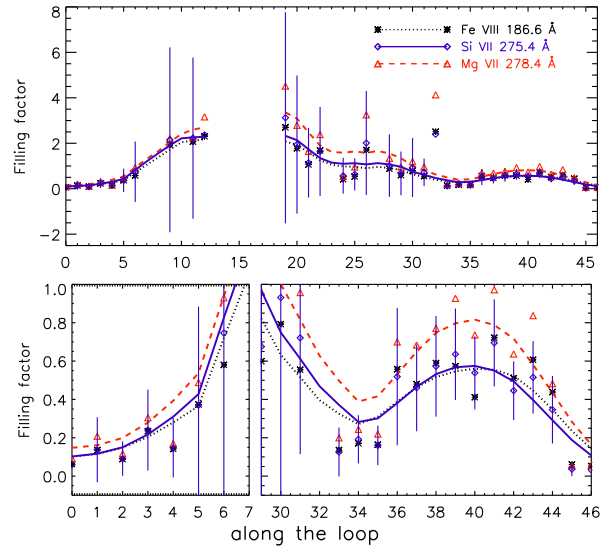}
\caption{Plasma filling factor ($\phi$) along the coronal loop from left foot-point A to right foot-point B as measured in different spectral lines.
5-point running average of data points are also over-plotted.}
\label{fig:loop_ff}
\end{figure}

To obtain the plasma filling factors ($\phi$) from various spectral lines using Equation~\ref{eq:phi}, we used the densities obtained from the 
\ion{Mg}{7} line ratio along the loop length, average loop width obtained from different spectral lines, and EM value (from Equation~\ref{eq:em})
is used for the corresponding loop temperature at that location as found in Figure~\ref{fig:temp_loci}. The plasma filling factors obtained from 
different spectral lines along the loop length are plotted in Figure~\ref{fig:loop_ff}. Error-bars on filling factors were obtained from standard 
procedures  which include errors on density, EM, and loop width (1-$\sigma$ error on FWHM) measurements. As it is evident from the plot
shown in the top panel of Figure~\ref{fig:loop_ff}, the filling factor values goes to 2, which is unrealistic. This is mainly the case near the loop top regions
where the errors on estimation of the column depth as well the determination of electron density is rather large due to very small signal to noise ratio.
In the bottom two plots of Figure~\ref{fig:loop_ff}, we have taken the left segment and right segment of the loop separately till the height where the filling
factor measurements could be considered real. These plots suggest that the filling factor increases as one moves from the loop foot-point to the loop top. 
Filling factors at the left foot-point A are about $0.11$ (11\%), $0.11$ (11\%), and  $0.16$ 
(16\%) as measured using the intensities from the \ion{Fe}{8}, \ion{Si}{7}, and \ion{Mg}{7} spectral lines respectively. The filling factor again decreases in 
all the spectral lines while approaching towards the right  foot-point B. However, it is observed that near the foot-point B, there are local enhancement in 
filling factors in all the spectral lines, this might have resulted from the appearance of other structures along the line-of-sight (which result in an under-estimation 
of column depth), as also evident from the intensity images obtained from EIS rasters as well as AIA images (see Figure~\ref{fig:loop_tracing}). 

The increase in filling factor values along the loop until the height it can be reliably measured is probably a consequence of the decrease
in density (it depends on the square of the density, Equation~\ref{eq:phi}) and can be considered real within the given error-bars. 
Note that the filling factor values obtained here are lower than those reported by \cite{2012ApJ...755L..33B} 
(for the loops having temperature about $\log\,T=6.2$), especially towards the foot-point regions.  
As mentioned earlier, the loop geometry may not be cylindrical, thus, our assumption of cylindrical geometry may have some effect
on the filling factor measurement. However, within the obtained error-bars, this effect is likely to be very small.

\subsection{Comparison with hydrostatic equilibrium}

The unique observation and measurement of density and temperature along a complete coronal loop made it possible to compare the loop
parameters with hydrostatic equilibrium. This has been done in past with the observations recorded using the Transition Region and 
Coronal Explorer \citep[TRACE; ][e.g. ]{2000ApJ...541.1059A}, but has never been done using spectroscopic observations. 

The electron density profile for a loop in hydrostatic equilibrium is given by:

%%%%------------------------------------------------------------
\begin{equation}
 n_e(h)=n_e(0)exp \left(- \frac{h}{\lambda (T_e)} \right)
\label{eq:hydrostatic}
\end{equation}
%%%%------------------------------------------------------------

where $\lambda$ is the density scale height given by,

%%%%------------------------------------------------------------
\begin{equation}
 \lambda(T_e)= \frac{k_b T_e}{\mu m_H g} \approx 46 \left[ \frac{T_e}{1~MK} \right] [Mm]
\label{eq:sheight}
\end{equation}
%%%%------------------------------------------------------------

Where $k_b$ is the Boltzmann constant, $T_e$ is electron temperature, $\mu$ is the mean molecular weight
 ($\approx$ 1.4 for the solar corona), $m_H$ is mass of the hydrogen atom, and $g$ is the acceleration due
 to gravity at the solar surface \citep[see e.g.,][]{1999ApJ...515..842A}. 

\begin{figure}[htbp]
\centering
\includegraphics[width=0.45\textwidth]{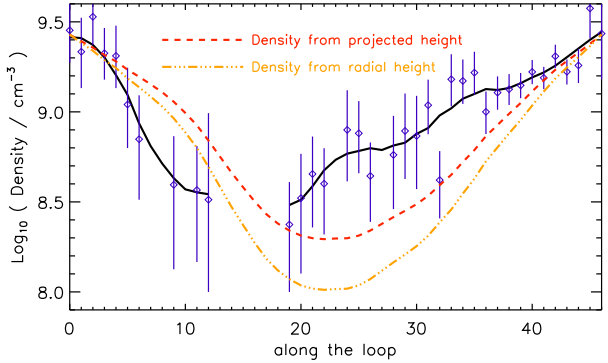}
\caption{Comparison of density as measured using Mg~{\sc vii} $\lambda 280.75/\lambda 278.39$ line ratio (plotted with diamond symbol and 5-point running average with continuous line)
from left foot-point A to right foot-point B with density expected from a hydrostatic model at temperature 0.73 MK obtained from projected (dashed line) and radial (dot-dashed line) heights of the loop as
found in Figure~\ref{fig:loop_height}.  }
\label{fig:hydro}
\end{figure}

Figure~\ref{fig:hydro} displays that variation of electron density with height as obtained from \ion{Mg}{7} line pair with diamonds 
and the solid black line is five points running average of the densities. We compare the density variation with height in the loop as 
obtained from Mg~{\sc vii} line pair with an exponentially decreasing density profile as given by Equation~\ref{eq:hydrostatic}.  
The density scale height was calculated for an electron temperature of 0.73 MK as obtained from the EM-loci method. We compare the density
variation with the projected (dashed line) and radially directed loop heights (dashed-dotted-dotted line)  as shown in Figure~\ref{fig:hydro}. Figure~\ref{fig:hydro}
 indicates that the observed density does not fall off with height as expected from the hydrostatic model. 
Although the density decreases with height, the fall is different to the expected hydrostatic equilibrium model. 
 Comparison shows that left half of loop is under-dense whereas right half of the loop is over-dense.
This result will also be hold true in the case of loop being inclined away or towards the observer with respect to solar-XY plane within the angle of $36\degree$. 
Thus, we suggest that the observed loop has non-symmetric density distribution as compared to the hydrostatic equilibrium and iso-thermal model.

%%%%------------------------------------------------------------
%%%%------------------------------------------------------------

\section{Discussions and summary} \label{conclusion}

Using the EIS/Hinode observations, we have studied the basic physical plasma parameters such as electron density, column depth, filling 
factors, temperature, and thermal pressure along a coronal loop from one foot-point to another. So far, the studies regarding coronal loops 
using spectrometers have been performed on small loop segments  \citep[see e.g.,][]{2001ApJ...556..896S,2003A&A...406.1089D,2007ApJ...655..598C,2008ApJ...686L.131W,2009ApJ...694.1256T}.
The measurements reported here make this paper unique as these parameters are derived from one foot-point to the another. Moreover, these measurements have allowed us to
perform a comparison, for the first time, of the observed loop properties with hydrostatic equilibrium obtained using spectroscopic data.

The results obtained are summarized below:
\begin{itemize}
\item Using the spectral line ratios of  \ion{Mg}{7} $\lambda 278.39/\lambda 280.75$, we derived electron 
densities along the loop length. The result indicates a fall of density from the loop foot-point to the loop top,  
the density measured at the loop foot-points is about $ 10^{9.4}$~cm$^{-3}$ which decreases to about $ 10^{8.5}$~cm$^{-3}$
towards the loop top.
 
 \item  We obtained the temperature of the observed loop, both  across (along the line-of-sight) and along the loop length using the EM-loci method.
 We found that the loop is almost iso-thermal across and along the loop with a temperature of about 0.73~MK. 
  
\item We measured the diameter of the loop along its length using the images obtained in different spectral lines and found that the loop has a non-uniform 
cross-section \citep[see Figure~\ref{fig:cdepth}, see also ][]{2014ApJ...796...20D}. We also estimated filling factors along the loop length and found that values obtained are similar to those reported 
previously. The filling factor obtained using  \ion{Mg}{7} 278.4~\AA\ , 
\ion{Si}{7} 275.4~\AA , and \ion{Fe}{8} 186.6~\AA\ spectral lines are similar to each other $\sim$~10\% (which increases with height) 
within the error bars. This is not surprising as the peak formation temperature for all these three lines is the same. Therefore, it is plausible to 
infer that the loop is multi-stranded.
 
\item The obtained electron density along the coronal loop from the \ion{Mg}{7} line pair is compared with the hydrostatic equilibrium
model at an iso-thermal temperature of 0.73 MK as was obtained from EM-loci study. We found that the observed densities in the left segment of 
the loop were lower (i.e. under-dense) than that expected from iso-thermal hydrostatic model. However, it was higher (i.e. over-dense) for the right 
segment of the loop (see Figure~\ref{fig:hydro}). This suggests a non-symmetric density profile along the loop.  However, we emphasize that this 
result is obtained under the assumption that observed loop is extending radially outward and will also hold true if the loop is inclined not only towards 
but also away from the observer with respect to solar-XY plane within the angle of $36\degree$ .
\end{itemize}

The results obtained here for the complete loop provides an opportunity for comparison with various loop models. 
The near iso-thermal nature of the loop along with the small filling factor and over density (super-hydrostatic)  is in agreement
with impulsive heating model   \citep{2004ApJ...605..911C,2006SoPh..234...41K}. However, the observations presented here shows the over density in only one part of the loop.
The quasi-steady foot-point heating model which drives the thermal non-equilibrium solutions, however, may explain the observed properties of the loop studies 
here \citep[e.g.][]{2013ApJ...773..134L, 2013ApJ...773...94M}, however, see \cite{2010ApJ...714.1239K,2013ApJ...779....1T}, and \cite{2014ApJ...791...60K}. 
Therefore, it will be interesting and important to measure physical properties such as densities, flows and geometries of many many well observed coronal loops 
and compare their properties with various loop heating models to make a generic conclusive statements on their heating and dynamics.

\acknowledgments
We acknowledge useful discussions at the International Space Science Institute (ISSI) workshops on \textquoteleft Active Region Heating\textquoteright .
We thank referee for his/her helpful comments and suggestions which have improved the quality of presentation.
G.R.G. is supported through the INSPIRE Faculty Fellowship of the Department of Science and Technology (DST), India.
H.E.M. acknowledges financial support from STFC (UK). 
Hinode is a Japanese mission developed and launched by ISAS/JAXA, collaborating with NAOJ as a domestic partner, 
NASA and STFC (UK) as international partners. Scientific operation of the Hinode mission is conducted by the Hinode science team
organized at ISAS/JAXA. This team mainly consists of scientists from institutes in the partner countries. Support for the post-launch 
operation is provided by JAXA and NAOJ (Japan), STFC (U.K.), NASA (U.S.A.), ESA, and NSC (Norway). 
The data used here are courtesy of NASA/SDO and AIA consortium.

\bibliographystyle{apj.bst}
\bibliography{references}

\end{document}